\documentclass[10pt,a4]{article}

\usepackage{epsfig}
\usepackage{amssymb}

\usepackage{graphics}

\begin{document}

\begin{titlepage}

\rightline{hep-ph/0503071}

\vskip 2cm

\centerline{\bf D-term inflation in D-brane cosmology}

\vskip 1cm

\centerline{G.Panotopoulos$^1$}

\vskip 1cm

\centerline{$^1$ Department of Physics, University of Crete,}

\vskip 0.2 cm

\centerline{Heraklion, Crete, Hellas}

\vskip 0.2cm

\centerline{email: {\it panotop@physics.uoc.gr}}

\begin{abstract}
We consider hybrid inflation in the braneworld scenario. In particular, we consider  inflation in global supersymmetry with the D-terms in the scalar potential for the inflaton field to be the dominant ones (D-term inflation). We find that D-term dominated inflation can naturally accomodate all requirements of the successful hybrid inflationary model also in the framework of D-brane cosmology with global supersymmetry. The reheating temperature after inflation can be high enough ($\sim 10^{10} \: GeV$) for successful thermal leptogenesis.  
\end{abstract}


\end{titlepage}

\section{Introduction}
Recently there has been considerable  interest in higher dimensional cosmological models. In those models our four-dimensional world lives on a three-dimensional extended object (brane) which is embedded in a higher dimensionl space (bulk). The models of this kind are  string-inspired ones, as it is known that in Type I string theory \cite{pol} there are two sectors, the open and the closed ones, and that the theory contains extended objects, called D-branes, where open strings can end. The fields in the closed sector (including gravity) can propagate in the bulk, whereas the fields in the open sector are confined to the brane. In such string-inspired scenarios the extra dimensions need not be small \cite{dvali} and in fact they can even be non-compact \cite{rs}. It is important to note that in the context of extra dimensions and the braneworld idea one discovers  a generalized Friedmann equation, which is different from the usual Friedmann equation in conventional cosmology. This means that the rate of expansion of the universe in this novel cosmology is altered and accordingly the physics in the  early universe can be different from what we know already. So it would be very interesting to study the cosmological implications of these new ideas about extra dimensions and braneworlds. Perhaps the best laboratory for such a study is inflation \cite{linde}, which has become the standard paradigm in the Big-Bang cosmology and which is in favour after the recent discovery from WMAP satellite (see e.g \cite{wmap}) that the universe is almost flat. It is known that there is not  a theory for inflation yet. All we have is a big collection of inflationary models. The single-field models for inflation, such as 'new' \cite{linde1} or 'chaotic' \cite{linde2}, are characterized by the disadvantage that they require 'tiny' coupling constants in order to reproduce the observational data. This difficulty was overcome by Linde who proposed, in the context of non-supersymmetric GUTS, the hybrid inflationary scenario \cite{linde3}. It turns out that one can consider hybrid inflation in supersymmetric theories (for a review on supersymmetry and supergravity see \cite{nilles}) too. In fact, inflation looks more natural in supersymmetric theories rather in non-supersymmetric ones \cite{riotto}. In a supersymmetric theory, the tree-level potential is the sum of an F-term and a D-term. These two terms have rather different properties and in all inflationary models only one of them dominates \cite{sakel}. The case of F-term inflation (where F-terms dominate) was considered for the first time in \cite{wands}, while the case of D-term inflation (where D-terms dominate) was considered in \cite{halyo}. In fact, if one considers supergravity then D-term inflation looks more promising, since it avoids the problem associated with the inflaton mass \cite{halyo}. F-term inflation in braneworld was studied in \cite{chafik}. In the present note we discuss the implications of D-term inflation. 

Before proceeding our discussion, let us specify our setup. The braneworld modelthat we shall consider is the supersymmetric version of the RS II model (see e.g\cite{bagger}). However,the cosmological solution of this extended model is the same as that in the non-supersymmetric model, since Einstein's equations belong to the bosonic part. The only sourse in the bulk is a five-dimensional cosmological constant. There is matter confined to the brane and during inflation, which is the cosmological era we shall be interested in, this matter is dominated by ascalar field, called the inflaton field $\phi$. 

The paper consists of six sections of which this introduction is the first. We present D-term inflation in the second section and brane cosmology in the third. Our results for the inflationary dynamics on the brane are discussed in the fourth section. We discuss reheating after inflation in the fifth section and finally we conclude in the fifth section.

\section{D-term inflation}
In this section we explain what D-term inflation is, following essentially \cite{riotto}. Inflation, by definition, breaks global supersymmetry since it requires a non-zero cosmological constant $V$ (false vacuum energy of the inflaton). For a D-term spontaneous breaking of supersymmetry a term linear in the auxiliary field $D$ is needed (Fayet-Iliopoulos mechanism \cite{fayet}). If the theory contains an abelian $U(1)$ gauge symmetry (anomalous or not), the Fayet-Iliopoulos D-term
\begin{equation}
\xi \int d^4\theta V=\xi D
\end{equation}
where $V$ is the vector superfield, is supersymmetric and gauge invariant and therefore allowed by the symmetries. We remark that an anomalous $U(1)$ symmetry is usually present in string theories and the anomaly is cancelled by the Green-Schwarz mechanism. However, here we will consider a non-anomalous $U(1)$ gauge symmetry.
In the context of global supersymmetry, D-term inflation is derived from the superpotential
\begin{equation}
W=\lambda \Phi \Phi_{+} \Phi_{-}
\end{equation}
where $\Phi, \Phi_{-}, \Phi_{+}$ are three chiral superfields and $\lambda$ is the superpotential coupling.  Under the $U(1)$ gauge symmetry the three chiral superfields have sharges $Q_{\Phi}=0, Q_{\Phi_{+}}=+1$ and $Q_{\Phi_{-}}=-1$, respectively. The superpotential given above leads to the following expression for the scalar potential
\begin{equation}
V(\phi_{+}, \phi_{-}, |\phi|)=\lambda^2 (|\phi|^2 (|\phi_{+}|^2+|\phi_{-}|^2)+|\phi_{+} \phi_{-}|^2)+\frac{g^2}{2} (|\phi_{+}|^2-|\phi_{-}|^2+\xi)^2
\end{equation}
where $\phi$ is the scalar component of the superfield $\Phi$, $\phi_{\pm}$ are the scalar components of the superfields $\Phi_{\pm}$, $g$ is the gauge coupling of the $U(1)$ symmetry and $\xi$ is a Fayet-Iliopoulos term, chosen to be positive. The global minimum is supersymmetry conserving, but the gauge group $U(1)$ is spontaneously broken
\begin{equation}
<\phi>=<\phi_{+}>=0, \; <\phi_{-}>=\sqrt{\xi}
\end{equation}
However, if we minimize the potential, for fixed values of $\phi$, with respect to other fields, we find that for $\phi>\phi_{c}=\frac{g}{\lambda} \sqrt{\xi}$, the minimum is at $\phi_{+}=\phi_{-}=0$. Thus, for $\phi>\phi_{c}$ and $\phi_{+}=\phi_{-}=0$ the tree-level potential has a vanishing curvature in the $\phi$ direction and large positive curvature in the remaining two directions $m_{\pm}^2=\lambda^2 |\phi|^2 \pm g^2 \xi$. 

For arbitrary large $\phi$ the tree-level value of the potential remains constant and equal to $V_{0}=(g^2/2) \xi^2$, thus $\phi$ plays naturally the role of an inflaton field. Along the inlationary trajectory the F-term vanishes and the universe is dominated by the D-term, which splits the masses in the $\Phi_{+}$ and $\Phi_{-}$ superfields, resulting to the one-loop effective potential for the inflaton field. The radiative corrections are given by the Coleman-Weinberg formula \cite{coleman}
\begin{equation}
\Delta V_{1-loop}=\frac{1}{64 \pi} \sum_{i} (-1)^{F_{i}} m_{i}^4 ln\frac{m_{i}^2}{\Lambda^2}
\end{equation}
where $\Lambda$ stands for a renormalization scale which does not affect physical quantities and the sum extends over all helicity states $i$, with fermion number $F_{i}$ and mass squared $m_{i}^2$.
The radiative corrections given by the above formula lead to the following effective potential for D-term inflation
\begin{equation}  \label{eq:1}
V(\phi)=\frac{g^2 \xi^2}{2} \left (1+\frac{g^2}{16 \pi^2} ln\frac{|\phi|^2 \lambda^2}{\Lambda^2} \right ) 
\end{equation}
The end of inflation is determined either by the failure of the slow-roll conditions or when $\phi$ approaches $\phi_{c}$.

\section{Effective gravitational equations on the brane}
Here we review the basic equations of brane cosmology. We work essentially in the context of Randall-Sundrum  II model \cite{rs}. In the bulk there is just a cosmological constant $\Lambda_{5}$, whereas on the brane there is matter with energy-momentum tensor $\tau_{\mu \nu}$. Also, the brane has a tension $T$. The five dimensional Planck mass is denoted by $M_{5}$.
If Einstein's equations hold in the five dimensional bulk, then it has been shown in \cite{shir} that the effective four-dimensionl Einstein's equations induced on the brane can be written as 
\begin{equation}
G_{\mu \nu}+\Lambda_{4} g_{\mu \nu}=\frac{8 \pi}{M_{p}^2} \tau_{\mu \nu}+(\frac{8 \pi}{M_{5}^3})^2 \pi_{\mu \nu}-E_{\mu \nu}
\end{equation}
where $g_{\mu \nu}$ is the induced metric on the brane, $\pi_{\mu \nu}=\frac{1}{12} \: \tau \: \tau_{\mu \nu}+\frac{1}{8} \: g_{\mu \nu} \: \tau_{\alpha \beta} \: \tau^{\alpha \beta}-\frac{1}{4} \: \tau_{\mu \alpha} \: \tau_{\nu}^{\alpha}-\frac{1}{24} \: \tau^2 \: g_{\mu \nu}$, $\Lambda_{4}$ is the effective four-dimensional cosmological constant, $M_{p}$ is the usual four-dimensional Planck mass and $E_{\mu \nu} \equiv C_{\beta \rho \sigma} ^\alpha \: n_{\alpha} \: n^{\rho} \: g_{\mu} ^{\beta} \: g_{\nu} ^{\sigma}$ is a projection of the five-dimensional Weyl tensor $C_{\alpha \beta \rho \sigma}$, where $n^{\alpha}$ is the unit vector normal to the brane.
The tensors $\pi_{\mu \nu}$ and $E_{\mu \nu}$ describe the influense of the bulk in brane dynamics. The five-dimensional quantities are related to the corresponding four-dimensional ones through the relations
\begin{equation}
M_{p}=\sqrt{\frac{3}{4 \pi}} \frac{M_{5}^3}{\sqrt{T}} 
\end{equation}
and
\begin{equation}
\Lambda_{4}=\frac{4 \pi}{M_{5}^3} \left( \Lambda_{5}+\frac{4 \pi T^2}{3 M_{5}^3} \right )
\end{equation}
In a cosmological model in which the induced metric on the brane $g_{\mu \nu}$ has the form of  a spatially flat Friedmann-Robertson-Walker model, with scale factor $a(t)$, the Friedmann-like equation on the brane has the generalized form \cite{binetry}
\begin{equation}
H^2=\frac{\Lambda_{4}}{3}+\frac{8 \pi}{3 M_{p}^2}  \rho+(\frac{4 \pi}{3 M_{5}^3})^2 \rho^2+\frac{C}{a^4}
\end{equation}
where $C$ is an integration constant arising from $E_{\mu \nu}$. The cosmological constant term and the term linear in $\rho$ are familiar from the four-dimensional convensional cosmology. The extra terms, i.e the ``dark radiation'' term and the term quadratic in $\rho$, are there because of the presense of the extra dimension. Adopting the Randall-Sundrum fine-tuning
\begin{equation}
\Lambda_{5}=-\frac{4 \pi T^2}{3 M_{5}^3}
\end{equation}
the four-dimensional cosmological constant vanishes. Furthermore, the term with the integration constant $C$ will be rapidly diluted during inflation and can be ignored. So the generalized Friedmann equation takes the final form
\begin{equation}
H^2=\frac{8 \pi}{3 M_{p}^2} \rho \left( 1+\frac{\rho}{2 T} \right )
\end{equation}
We notice that in the low density regime $\rho \ll T$ we recover the usual Friedmann equation. However, in the high energy regime $\rho \gg T$ the unity can be neglected and then the Friedmann-like equation becomes
\begin{equation}
H^2=\frac{4 \pi \rho^2}{3 T M_{4}^2}
\end{equation}
Note that in this regime the Hubble parameter is linear in $\rho$, while in conventional cosmology it goes with the square root of $\rho$.

\section{Inflationary dynamics on the brane}
As already mentioned, we will consider the case in which the energy momentum on the brane is dominated by a scalar field $\phi$ confined on the brane with a self-interaction potential $V(\phi)$ given in (\ref{eq:1}). The field $\phi$ is a function of time only, as dictated by the isotropy and homogeneity of the observed four-dimensional universe. A homogeneous scalar field behaves like a perfect fluid with pressure $p=(1/2) \dot{\phi}^2-V$ and energy density $\rho=(1/2) \dot{\phi}^2+V$. There is no energy exchange between the brane and the bulk, so the energy-momentum tensor $T_{\mu \nu}$ of the scalar field is conserved, that is $\nabla ^ \nu T_{\mu \nu}=0$. This is equivalent to the continuity equation for the pressure $p$ and the energy density $\rho$
\begin{equation}
\dot{\rho}+3 H (p+\rho)=0
\end{equation}
where $H$ is the Hubble parameter $H=\dot{a}/a$. Therefore we get the equation of motion for the scalar field $\phi$, which is the following
\begin{equation}
\ddot{\phi}+3 H \dot{\phi}+V'(\phi)=0
\end{equation}
This is of course the Klein-Gordon equation for a scalar field in a Robertson-Walker background. The equation that governs the dynamics of the expansion of the universe is the Friedmann-like equation of the previous section. Inflation takes place in the early stages of the evolution of the universe, so in the Friedmann equation the extra term dominates and therefore the equation for the scale factor is
\begin{equation}
H^2=\frac{4 \pi \rho^2}{3 T M_{p}^2}
\end{equation}
In the slow-roll approximation the slope and the curvature of the potential must satisfy the two constraints $\epsilon \ll 1$ and $|\eta| \ll 1$, where $\epsilon$ and $\eta$ are the two slow-roll parameters which are defined by
\begin{equation}
\epsilon \equiv -\frac{\dot{H}}{H^2}
\end{equation}
\begin{equation}
\eta \equiv \frac{V''}{3 H^2}
\end{equation}
In this approximation the equation of motion for the scalar field takes the form
\begin{equation}
\dot{\phi} \simeq -\frac{V'}{3 H}
\end{equation}
while the generalized Friedmann equation becomes $(V \gg \dot{\phi}^2)$
\begin{equation}
H^2 \simeq \frac{4 \pi V^2}{3 T M_{p}^2}
\end{equation}
The number of e-folds during inflation is given by
\begin{equation}
N \equiv ln \frac{a_{f}}{a_{i}} = \int _{t_{i}}^{t_{f}} \: H dt
\end{equation}
For a strong enough inflation we take $N=60$.
In the slow-roll approximation the number of e-folds and the slow-roll parameters are given by the formulae \cite{maartens}
\begin{equation}
\epsilon \simeq \frac{M_{p}^2}{16 \pi} \: \left (\frac{V'}{V} \right )^2 \: \frac{4 T}{V}
\end{equation}
\begin{equation}
\eta \simeq \frac{M_{p}^2}{8 \pi} \: \left (\frac{V''}{V} \right )^2  \: \frac{2 T}{V}
\end{equation}
\begin{equation}
N \simeq -\frac{8 \pi}{M_{p}^2} \: \int _{\phi_{i}} ^ {\phi_{f}} \: \frac{V}{V'} \: \frac{V}{2 T} \: d \phi
\end{equation} 
The main cosmological constraint comes from the amplitude of the scalar perturbations which is given in this new context by \cite{maartens}
\begin{equation}
A_{s}^2=\frac{512 \pi}{75 M_{p}^6} \: \frac{V^3}{V'^2} \: \left (\frac{V}{2 T} \right )^2 
\end{equation}
where the right-hand side is evaluated at the horizon-crossing when the comoving scale equals the Hubble radius during inflation. Finally, the spectral index for the scalar perturbations is given in terms of the slow-roll parameters
\begin{equation}
n_{s}-1 \equiv \frac{d \: ln A_{s}^2}{d \: ln k}=2 \eta - 6 \epsilon
\end{equation}
and the tensor-to-scalar ratio is given by
\begin{equation}
\frac{A_{t}^2}{A_{s}^2}=\epsilon \: \frac{T}{V}
\end{equation}
In what follows we will assume that $g \sim 0.5$ and that inflation ends at $\phi_{c}=(g/ \lambda) \: \sqrt{\xi}$. To make sure that the slow-roll conditions are satisfied we impose the constraint
\begin{equation}
\frac{T M_{p}^2 \lambda^2}{16 \pi^3 g^2 \xi^3} \ll 1
\end{equation}
Also, we have assumed that the potential $V$ is much larger than the brane tension $T$. Therefore another constraint to be satisfied is
\begin{equation}
\frac{g^2 \xi^2}{4 T} \gg 1
\end{equation}
Now that we have written all the necessary formulae, we can proceed to the presentation of our results. For arbitrary $\lambda$ it is not possible to satisfy both the datum from COBE that $A_{s}=2 \cdot 10^{-5}$ and the slow-roll conditions. For this to happen the superpotential coupling $\lambda$ has to be smaller or equal to $0.0185$ (approximately). Then, for a given value for $\lambda$, the brane tension cannot become arbitrarily large because in that case the constraint that the potential should be much larger than the brane tension is not satisfied. We find the following upper bound for the brane tension $T$
\begin{equation}
T \leq 5 \cdot 10^{55} \: GeV^4
\end{equation}
Now that we have set upper bounds for $T$ and $\lambda$ so that our constraints and the data from COBE are satisfied, we can compute the spectral index $n_{s}$ and the tensor-to-scalar ratio $r$.
For example, for the values $T=5 \cdot 10^{55} \: GeV^4$ and $\lambda=0.0185$ we find
\begin{equation}
n_{s}=0.99, \;  r=2 \cdot 10^{-4}
\end{equation}
A detailed analysis shows that for a particular value for $\lambda$ (below the upper bound of course) the spectral index does not depend on $T$ and is always very close to 1. As $\lambda$ becomes smaller and smaller the spectral index slightly increases and gets even closer to 1. Also, in all cases the tensor perturbations are negligible.
Finally, we find that for the maximum value for the brane tension $\sqrt{\xi} \sim 10^{14} \: GeV$, whereas $\sqrt{\xi}$ becomes smaller as $T$ decreases. We note that according to our analysis $\lambda$ a priori can take arbitrarily small values. However, this would be unnatural and for that reason we do not consider values for $\lambda$ much smaller than $10^{-3}$. In that case we find that the values of the inflaton remain safely below Planck mass and therefore global supersymmetry is a good approximation.

\section{Reheating}
Finally, let us turn to the discussion of reheating after inflation and to the computation of the reheating temperature $T_R$. After slow-roll the inflaton decays with a decay rate $\Gamma$ and the decay products quickly thermalize. This is the way the universe re-enters the radiation era of standard Big-Bang cosmology. The reheating temperature $T_R$ is related to two more cosmological topics, namely the gravitino problem \cite{khlopov} and the baryogenesis through leptogenesis. In gravity mediated SUSY breaking models and for an interesting range of the gravitino mass, $m_{3/2} \sim 0.1-1 TeV$, if the gravitino is unstable it has a long lifetime and decays after the BBN. The decay products destroy light elements produced by the BBN and hence the primordial abundance of the gravitino is constrained from above to keep the success of the BBN. This leads to an upper bound on the reheating temperature $T_R$ after inflation, since the abundance of the gravitino is proportional to $T_R$. A detailed analysis derived a stringent upper bound $T_R \leqslant 10^6-10^7 \: GeV$ when gravitino has hadronic modes \cite{ellis}. On the other hand, primordial lepton asymetry is converted to baryon asymmetry \cite{yanagida} in the early universe through the ``sphaleron'' effects of the electroweak gauge theory \cite{rubakov}. This baryogenesis through leptogenesis requires a lower bound on the reheating temperature. Leptogenesis can be thermal or non-thermal. For a thermal leptogenesis $T_{R} \geqslant 2 \cdot 10^9 \: GeV$ \cite{giudice}, whereas for non-thermal leptogenesis $T_{R} \geqslant 10^6 \: GeV$ \cite{asaka}. It seems that it is impossible to satisfy both constaints for the reheating temperature coming from leptogenesis and the gravitino problem. However, the authors of \cite{gravitino} have showed that in the brane world scenario, that we discuss here, it is possible to solve the gravitino problem allowing for the reheating temperature to be as high as $10^{10} \: GeV$. According to ref. \cite{gravitino} the gravitino abundance is proportional not to the reheating temperature, as is the case in conventional cosmology, but to a transition temperature $T_t$ between high temperatures ($T_R$) and low ones (today's temperature $T_0$). That way the requirement for not over-production of gravitini leads to an upper bound for this transition temperature and not for the reheating temperature, which can be as high as a satisfactory leptogenesis requires. 

The reheating temperature is given by the formula
\begin{equation}
T_R=\left (\frac{90 T \Gamma^2 m_p^2}{2 \pi^3 g^2 \xi^2 g_{eff}^2} \right )^{1/4}
\end{equation}
where $g_{eff}$ is the effective number of degrees of freedom at the reheating temperature and for the MSSM is $g_{eff}=\frac{915}{4}$. Assuming that the inflaton $\phi$ decays to the lighest of the three heavy right handed neutrinos $\psi$ 
\begin{equation}
\phi \; \rightarrow \; \psi + \psi
\end{equation}
the decay rate of the inflaton is \cite{sakel}
\begin{equation}
\Gamma=\frac{m_{infl}}{8 \pi} \left ( \frac{M_1}{\sqrt{\xi}} \right )^2
\end{equation}
where $m_{infl}$ is the inflaton mass, $M_1$ is the smallest of the three neutrino mass eigenvalues and $m_{infl} > 2 M_{1}$.  The mass of the inflaton is given in terms of the coupling constant $g$ and the Fayet-Iliopoulos parameter $\xi$ by
\begin{equation}
m_{infl} = \sqrt{2} \: g \: \sqrt{\xi}
\end{equation} 
If the value of the mass of the lightest right handed neutrino is $M_{1}=10^{10} \: GeV$, which is a representative value, then the reheating temperature $T_{R}$ turns out to be of the order of the right handed neutrino mass. In fact the reheating temperature is a little bit larger than the neutrino mass. So we see that the reheating temperature is of the right order of magnitude for thermal leptogenesis. When the right handed neutrino mass increases (remaining though smaller than $m_{infl}/2$), the reheating temperature increases too and is always of the order of the neutrino mass but a little bit larger (see Figure 1). For a given value of $M_{1}$ and of the superpotential coupling $\lambda$ the reheating temperature does not change varying the tension of the brane $T$. Finally, for a given $M_1$, when $\lambda$ increases then $T_{R}$ decreases, but only slighty so as to remain of the order of magnitude of the mass $M_{1}$ (see Figure 2).

\section{Conclusions}
To summarize,  we have reexamined supersymmetric D-term dominated hybrid inflation in brane cosmology. We have found that we can reproduce the observational data provided that each of the brane tension, five-dimensional Planck mass and the superpotential coupling does not exceed a particular value. For a given value for the superpotential coupling, when the brane tension takes the maximum allowed value then the scale of inflation $\sqrt{\xi}$ is of the order of $\sim 10^{14} \: GeV$. This value of the inflationary scale is lower than the (supersymmetric) GUT scale, but close to it. Also, we have found that for natural values of the superpotential coupling $\lambda$ the inflaton field cannot take large values and stays well below the four-dimensional Planck mass, consistent with the global supersymmetry approximation adopted here. Furthermore, we have seen that our results are compatible with the corresponding results in the standard four-dimensional cosmology. This means that the advantages of the hybrid model are naturally preserved in the framework of brane cosmology. Finally, our study shows that the reheating temperature after inflation can naturally be of order $10^{10} \: GeV$ (or larger) allowing for a successful  thermal leptogenesis. 

\section*{Acknowlegements}

The author would like to thank T.N.Tomaras for valuable comments on the manuscript and I.Antoniadis and G.Kofinas for useful discussions. The author would like also to thank CERN theory division (where part of this work was completed) for its warmest hospitality. Work supported by the Greek Ministry of education research program "Heraklitos" and by the EU grant MRTN-CT-2004-512194.

\newpage

\begin{figure}
\centerline{\epsfig{figure=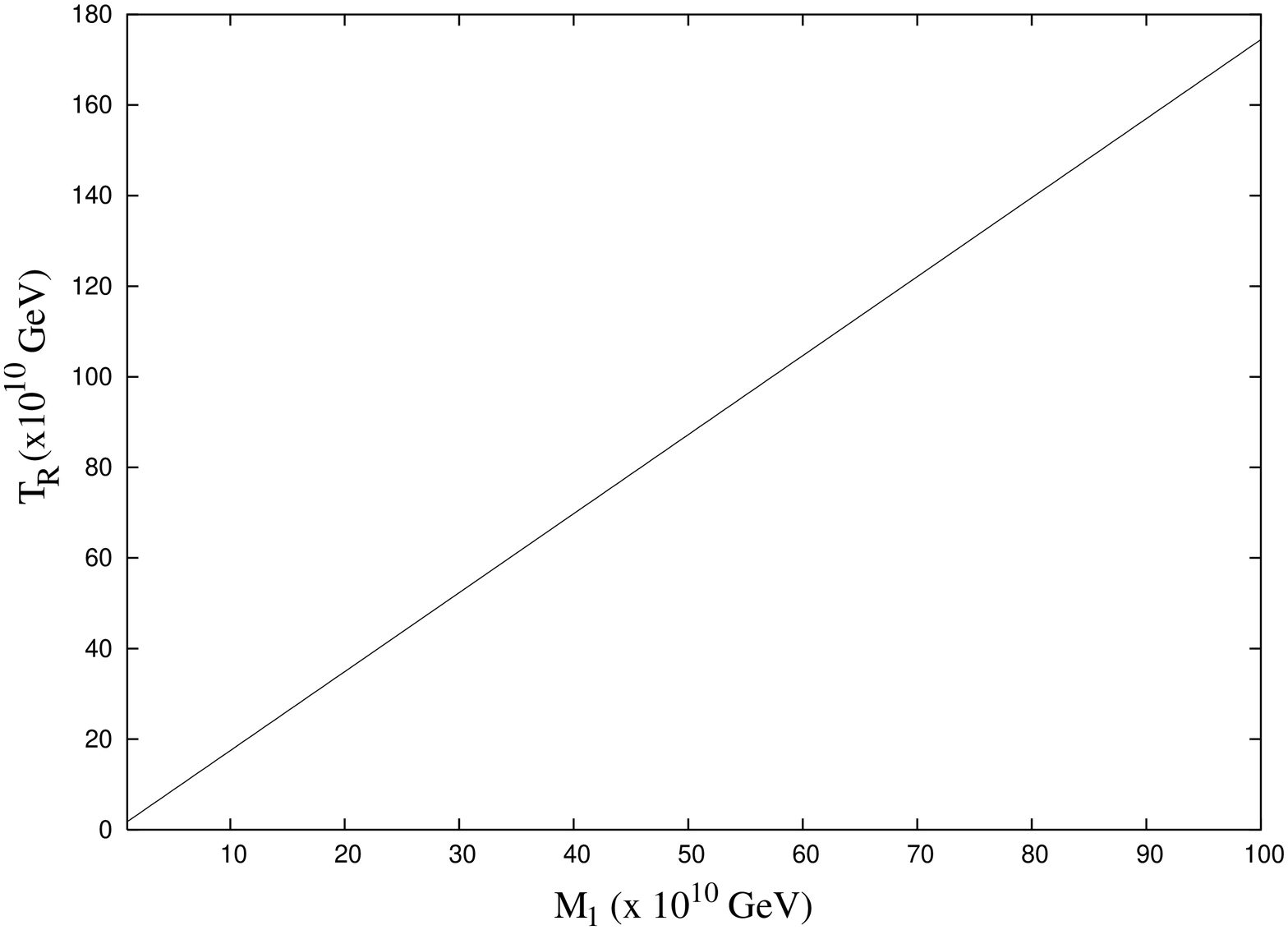,height=14cm,angle=0}}
\caption{Reheating temperature $T_R$ versus the right-handed neutrino mass $M_1$ for superpotential coupling $\lambda=0.001$.}
\end{figure}

\newpage

\begin{figure}
\centerline{\epsfig{figure=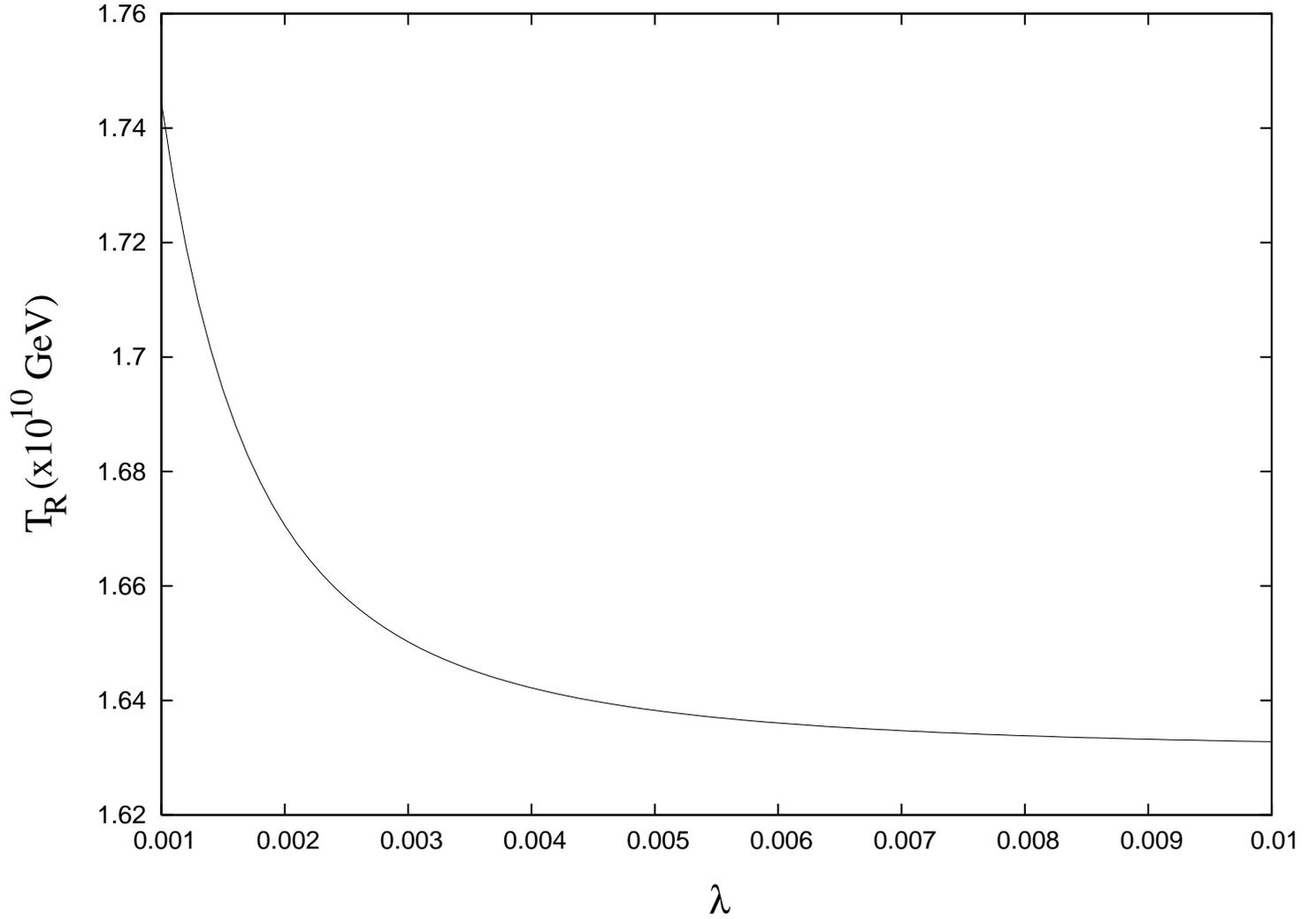,height=14cm,angle=0}}
\caption{Reheating temperature $T_R$ versus the superpotential coupling $\lambda$ for $M_1=10^{10} \: GeV$.}
\end{figure}

\end{document}